# MAGNETIC FREEZING EFFECT FOR THE GROUND STATE OF QUANTUM DOT

**V.D. Krevchik**[1], **A.B. Grunin**[1], **A.K. Aringazin**[2,3], and **M.B. Semenov**[1,3]

[1]Department of Physics, Penza State University, Penza, 440017, Russia
physics@diamond.stup.ac.ru
[2]Department of Theoretical Physics, Institute for Basic Research,
Eurasian National University, Astana, 473021, Kazakstan
[3]Institute for Basic Research, P.O. Box 1577, Palm Harbor, Fl 34682, USA
ibr@gte.net

Within the framework of zero-range potential model in the approach of effective mass, the impurity absorption by the complex Quantum Dot – Impurity Center in an external constant uniform magnetic field is considered. Under condition that the influence of a magnetic field on the ground state of quantum dot is negligible, we have derived the light absorption coefficient of impurity for the case of longitudinal polarization. It is shown that with an increase of the intensity of magnetic field the threshold of an impurity absorption band is shifted to the short-wave spectrum region. Also, the absorption coefficient increases by several times that can be explained as a result of the "magnetic freezing" effect for the ground state of quantum dot.

# 1. Introduction

Magneto-optics for semiconductive quantum dots (QD), which are synthesized in a transparent dielectric matrix, is of great interest for the research of fundamental properties of quasi-zero-dimensional structures and the influence of magnetic fields on optical transitions. An energy dependence of optical transitions both on the quantum dot size (QD size) and magnetic field intensity opens up possibilities for various applications in opto-electronics. Moreover, experiments on the photoluminescence spectrum dynamics in semiconductive heterostructures with quantum dots in an external magnetic field show [1, 2] that transition energies can be effectively changed by comparatively weak magnetic fields ranging from 4 to 12 Tesla.

The $\delta$-doping technology [3] allows us to consider an additional parameter – the impurity level depth – which influences on optical transitions in quantum dot.

In this paper, we consider magneto-optics for the Quantum Dot – Impurity Center (**QD-IC**) complexes, which are synthesized in a transparent dielectric matrix. The spherically symmetric oscillatory potential model is used to describe QD, and the impurity potential is simulated by the zero-range potential [4-7]. Within the approach of effective mass, we calculate the light impurity absorption coefficient in the presence of an external constant uniform magnetic field with an account of the dispersion of quantum dot sizes.

It is shown that an external magnetic field allows one to change not only the position of the band edge but also the light impurity absorption value. Also it is shown that under these conditions the light impurity absorption edge position essentially depends on QD parameters and the impurity level depth.

# 2. The absorption coefficient

We consider the light absorption by QD-IC complex in an external magnetic field $\vec{B}$. The field is assumed to be relatively weak so that its influence on the ground impurity state in QD is negligibly small. Evidently, this holds[1)] if $E_\lambda >> \hbar\Omega$, where $\Omega = |e| B / m^*$ is cyclotron frequency; $m^*$ is effective mass of electron, $|e|$ is charge of electron, and $E_\lambda = -\hbar^2\lambda^2/2m^*$ is the impurity ground state energy.

Within the framework of zero-range potential model, the following wave function describes the ground state for impurity center, which has been localized in point with $\vec{R}_a$=(0,0,0) [10]:

$$\Psi_\lambda(r) = \left\{ \frac{2\sqrt{\pi}\Gamma(\beta/2+1)a^3}{\beta^2\Gamma(\beta/2-1/2)} \left[ \frac{\beta}{2}\left(\Psi(\beta/2+1) - \Psi(\beta/2-1/2)\right) - 1 \right] \right\}^{-1/2} \times$$

$$\times \left( \frac{r^2}{a^2} \right)^{-3/4} \cdot \Gamma\left(\frac{\beta}{2}\right) \cdot W_{-\frac{\beta}{2}+\frac{3}{4}, \frac{1}{4}} \left( \frac{r^2}{a^2} \right), \qquad (1)$$

where $W_{\kappa,\mu}(x)$ is Whittaker function, $\Gamma(x)$ is Euler gamma-function; $\Psi(x)$ is derivative of natural logarithm of the gamma-function; $a = \sqrt{\hbar / (m^* \omega_0)}$; $E_0 = 3/2 \cdot \hbar\omega_0$ is ground state energy of QD; $\beta = \eta^2 + 3/2$; $\eta = \sqrt{|E_\lambda| / E_d}$; $E_d = m^* e^4 / (32\hbar^2\pi^2\varepsilon_0^2\varepsilon^2)$ is effective Bohr energy, with the effective mass $m^*$ and the dielectric permeability $\varepsilon$; and $\varepsilon_0$ is dielectric constant. The parameter $\eta$ satisfies the following equation [10]:

$$\sqrt{\eta^2 + \frac{3}{2}\beta^{-1}} = \eta_i - \sqrt{\frac{2}{\beta\pi}} \int_0^\infty dt \, \exp\left[ -\left( \beta\,\eta^2 + 3/2 \right) t \right] \times$$

$$\times \left[ \frac{1}{2t\sqrt{2t}} - \frac{1}{\left(1 - \exp(-2t)\right)^{3/2}} \cdot \exp\left\{ -\frac{R_a^{*2}\beta^{-1}}{2} \frac{\left[1 - \exp(-t)\right]}{\left[1 + \exp(-t)\right]} \right\} \right], \qquad (2)$$

where $\eta_i = \sqrt{|E_i| / E_d}$ and $E_i$ is impurity ground state energy in massive semiconductor.

---

[1)] Photo-ionization of deep impurity centers in an external magnetic field for the case of massive semiconductor was investigated theoretically in Refs.[8, 9].

For the case of strong localization of the impurity electron, $\lambda a$>>1, the wave function of the final state, $\Psi_{n1,m,n2}(\rho, \varphi, z)$, in the presence of an external magnetic field (nonsymmetric gauge fixing of the vector-potential, $\vec{A} = \left[\vec{B},\vec{r}\right]/2$), and corresponding energies $E_{n1,m,n2}$ have the following form:

$$\Psi_{n_1,m,n_2}(\rho,\varphi,z) = \frac{1}{a_1^{|m|+1}}\left[\frac{(n_1+|m|)!}{2^{n_2+|m|+1} n_1! n_2! \pi\sqrt{\pi}(|m|!)^2 a}\right]^{1/2} \rho^{|m|} \times$$

$$\times \exp\left[-\left(\frac{\rho^2}{4a_1^2}+\frac{z^2}{2a^2}\right)\right] H_{n_2}\left(\frac{z}{a}\right) F\left(-n_1, |m|+1, \frac{\rho^2}{2a_1^2}\right)\exp(im\varphi), \quad (3)$$

$$E_{n_1,m,n_2} = -\frac{eB\hbar m}{2m^*} + \hbar\omega_0(n_2+1/2) + \hbar\sqrt{\omega_0^2+\frac{e^2B^2}{4m^{*2}}}(2n_1+|m|+1), \quad (4)$$

where $\rho, \varphi, z$ are cylindrical coordinates, $H_n(x)$ is Hermite polynomial; $F(\alpha,\beta,x)$ is confluent hypergeometric function; $a_1^2=a^2/(2\sqrt{1+a^4/4a^4_B})$; $a_B=\sqrt{\hbar/(m^*\Omega)}$; $n_1, n_2 = 0, 1, 2,\ldots$ are quantum numbers corresponding to Landau levels and to energy levels for spherically-symmetric oscillator potential; and $m = 0, \pm 1, \pm 2,\ldots$ is the magnetic quantum number.

In the case of longitudinal polarization, the effective Hamiltonian for interaction with the light has the form

$$H_{\text{int}} = \lambda_0\sqrt{\frac{2\pi\hbar^2\alpha^*}{m^{*2}\omega}I_0}\, e^{i\vec{q}\vec{r}}\, (\vec{e}_\lambda\, \hat{\vec{P}}), \quad (5)$$

where $\lambda_0$ is local field coefficient; $\alpha^*$ is the fine structure constant with account of dielectric permeability $\varepsilon$; $I_0$ is intensity, $\omega$ is frequency of the light; $q$ is wave vector; $\vec{e}_\lambda$ is unit vector of the longitudinal polarization.

The matrix element which determines oscillator force value for the electron's dipole optical transition from the ground state of impurity center to states of discrete QD spectrum (these states are described by the wave function (3)), can be obtained as

$$M_{f\lambda}=\frac{(-1)^n i\pi\lambda_0}{2^{n+1}n!a_1}\sqrt{\frac{\alpha^* I_0}{\omega}}\beta a^2\frac{[(2n+1)!]^{1/2}\Gamma\left(\frac{\beta}{2}+n\right)\left[\Gamma\left(\frac{\beta}{2}-\frac{1}{2}\right)\right]^{1/2}}{\left[\Gamma\left(\frac{\beta}{2}+1\right)\right]^{1/2}\left[\frac{\beta}{2}\left(\Psi\left(\frac{\beta}{2}+1\right)-\Psi\left(\frac{\beta}{2}-\frac{1}{2}\right)\right)-1\right]^{1/2}}\times$$

$$\times\left(E_{n_1,0,2n+1}-E_\lambda\right)\sum_{k=0}^{n_1}(-1)^k C_{n_1}^k\left(1+\frac{a^4}{4a_B^4}\right)^{k/2}\frac{2^{k+1}\Gamma(k+2)}{\left(1+\sqrt{1+\frac{a^4}{4a_B^4}}\right)^{k+1}}\times$$

$$\times\frac{1}{\Gamma\left(\frac{\beta}{2}+n+k+2\right)}F\left(\frac{\beta}{2}+n,k+1;\frac{\beta}{2}+n+k+2,1-\frac{2}{1+\sqrt{1+\frac{a^4}{4a_B^4}}}\right),\quad(6)$$

where $F(\alpha,\beta;\gamma,z)$ is hypergeometric Gauss function. The formula (6) takes into account selection rules, which can be derived due to the integral,

$$\int_{-\infty}^{\infty}u\exp\left[-(1+t)u^2\right]H_{n_2}(u)du=\begin{cases}0,\ \text{if } n_2\neq 2n+1,\ n=0,1,2,...,\\(1+t)^{-3/2-n}\sqrt{\pi}\frac{(2n+1)!}{n!}(-1)^n t^n,\ \text{if } n_2=2n+1.\end{cases}\quad(7)$$

Eq. (7) shows that in the case of longitudinal polarization (with respect to the direction of an external magnetic field) optical transitions from impurity level are possible only to the states with odd values of the quantum number $n_2$.

Consideration of the quantum-dot-size dispersion[2)] leads to the following impurity absorption coefficient *K(ω)*:

$$K(\omega)=K_0\frac{\left(\eta^2+\frac{3}{2}\right)^2\Gamma\left(\frac{\eta^2}{2}+\frac{1}{4}\right)\beta^*X}{\Gamma\left(\frac{\eta^2}{2}+\frac{7}{4}\right)\left[\left(\frac{\eta^2}{2}+\frac{3}{4}\right)\left(\Psi\left(\frac{\eta^2}{2}+\frac{7}{4}\right)-\Psi\left(\frac{\eta^2}{2}+\frac{1}{4}\right)\right)-1\right]}\times$$

$$\times\sum_{n_1,n}\frac{(2n_1+1)(2n+1)!\Gamma^2\left(\frac{\eta^2}{2}+\frac{3}{4}+n\right)}{2^{2n+1}(n!)^2}\frac{u^{*4}_{n,n_1}\exp\left[-1/\left(1-2u^*_{n,n_1}/3\right)\right]}{\left(u^*_{n,n_1}+3\right)^{7/3}\left(3/2-u^*_{n,n_1}\right)^{11/3}}\times$$

$$\times\left|\beta^*u^*_{n,n_1}\left[\frac{(2n_1+1)^2}{a^{*4}}-\left(X-\eta^2\right)^2\right]+\left(2n+\frac{3}{2}\right)\left(X-\eta^2\right)\right|^{-1}\times$$

$$\times\left[\sum_{k=0}^{n_1}(-1)^kC^k_{n_1}\frac{2^{k+1}\Gamma(k+2)}{\Gamma\left(\frac{\eta^2}{2}+\frac{11}{4}+n+k\right)}\frac{\left[\beta^*\left(X-\eta^2\right)u^*_{n,n_1}-\left(2n+\frac{3}{2}\right)\right]^{k+1}}{\left[\beta^*\left(X-\eta^2\right)u^*_{n,n_1}+2(n_1-n)-\frac{1}{2}\right]^{k+1}}\times\right.$$

---

[2)] It is supposed that the dispersion arises during phase decay process in resaturated solid solution [11, 12] and has been satisfactorily described by Lifshits-Slezov formula [13],

$$P(u=R_0/\overline{R}_0)=\begin{cases}\dfrac{3eu^2\exp[-1/(1-2u/3)]}{2^{5/3}(3+u)^{7/3}(3/2-u)^{11/3}}, & u<3/2\\ 0, & u>3/2\end{cases}$$

where **e** is the natural logarithm base; $R_0$ и $\overline{R_0}$ are QD radius and mean value of QD radius, respectively.

$$\times F\left(\frac{\eta^2}{2}+\frac{3}{4}+n,k+1,\frac{\eta^2}{2}+\frac{11}{4}+n+k,1-\frac{2(2n_1+1)}{\beta^*\left(X-\eta^2\right)u^*_{n,n_1}+2(n_1-n)-\frac{1}{2}}\right)\Bigg]^2, \tag{8}$$

where $K_0=3\cdot 2^{1/3}\pi^3\alpha^*\lambda_0 e\, a_d^2 N_0$; $N_0$ is QD concentration in dielectric matrix; $X = \hbar\omega/E_d$ is photon energy in units of effective Bohr energy; $a_d$ is the effective Bohr radius, and $u^*_{n,n_1}$ is defined as

$$u^*_{n,n_1}=\frac{\left(2n+\frac{3}{2}\right)\left(X-\eta^2\right)+(2n_1+1)\sqrt{\left(X-\eta^2\right)^2+\frac{1}{a^{*4}}\left(\left(2n+\frac{3}{2}\right)^2-(2n_1+1)^2\right)}}{\beta^*\left(\left(X-\eta^2\right)^2-\frac{1}{a^{*4}}(2n_1+1)^2\right)}, \tag{9}$$

where $a^*= a_B /a_d$; $\beta^* = \overline{R}_0^* / \left(4\sqrt{U_0^*}\right)$; $U_0^*=U_0/E_d$; and $U_0$ is the amplitude of QD potential.

In Fig. 1, spectral dependences of the normalized light impurity absorption coefficient $K/K_0$ on the magnetic field intensity, for optical transition with maximal oscillator force ($n_1=n=0$), are plotted. It is evident that the consideration of the QD size dispersion gives a broadening of discrete lines in the absorption coefficient. With an increase of the intensity of magnetic field (i.e., with a decrease of the parameter $a^*=\sqrt{\hbar / \left(a_d^{\ 2} B\,|e|\right)}$), the threshold of an impurity absorption band is shifted to the short-wave spectrum region that is related to the corresponding dynamics of the Landau level. This shift is in accord to $X_t \approx \eta^2+8\sqrt{U_0^*}\left(3/2+\sqrt{1+9\overline{R}_0^{*2}/(64U_0^* a^{*4})}\right)/(3\overline{R}_0^*)$, where $\overline{R}^*_0 = 2\overline{R}_0 / a_d$. For higher intensities of the magnetic field, the light impurity absorption coefficient increases (compare curves 1 and 2 of Fig. 1). This effect can be qualitatively explained by an effective dimensional reduction of the spherically-symmetric oscillator potential. Indeed, since the inequality $a_B<a$ holds and is stronger for higher intensity magnetic fields, the confinement of the motion of charge carriers to the $x$-$y$ plain which occurs due to "weakening" of QD potential, can be viewed as a perturbation (compare curves 2 and 3 of Fig. 1). With such a reduction of the QD potential dimension the overlap between wave functions of initial and final states is deeper that, consequently, implies an essential increase of the optical transition probability.

The dependence of the impurity absorption threshold, $X_t = (\hbar\omega)_{threshold} / E_d$, on the parameter $a^*$ for different QD potential amplitude values, $U_0^* = U_0/E_d$, and zero-range potential intensities ($\eta_i$) is plotted in Fig. 2. In the weak-field region ($a^*>0.3$), $X_t$ does not depend essentially on the field intensity (horizontal parts of the curves 1 – 4 of Fig. 2) and mainly is determined by the impurity level depth and the QD potential amplitude. In the strong-field region ($a^*<0.3$), a considerable increase of $X_t$ takes place. This increase is related to the Landau level motion.

As an example, we take QD based on InSb which is characterized by $a_d \approx 750$ Å, and $E_d \approx 6\times10^{-4}$ eV. In this case, the shift of the impurity absorption band threshold is estimated to be about 0.05 eV, for the magnetic field intensity increase from 1.3 T to 12 T (or, from $a^*=0.3$ to $a^*=0.1$, respectively; see Fig. 2).

## 3. Summary

We have considered theoretical aspects of the impurity absorption by QD-IC complexes, which are synthesized in transparent dielectric matrix. The zero-range potential model has been used for the impurity potential, and QD has been described within the framework of a parabolic holding potential (QD amplitude $U_0^*$ is an empirical parameter here). The behavior of the impurity absorption spectrum with the variation of external magnetic field intensity has been investigated. As a result of the field intensity increase, the threshold of the impurity absorption band is shifted to the short-wave spectrum region, and the absorption coefficient value increases by several times. Since we consider $|E_\lambda| >> \hbar\Omega$ the increase of the impurity absorption coefficient (in the case of longitudinal polarization) in an external magnetic field can be viewed a result of the "magnetic freezing" effect for the QD ground state. The example of QD based on InSb shows that a quite effective control of the impurity absorption band can be done with the help of comparatively weak magnetic fields.

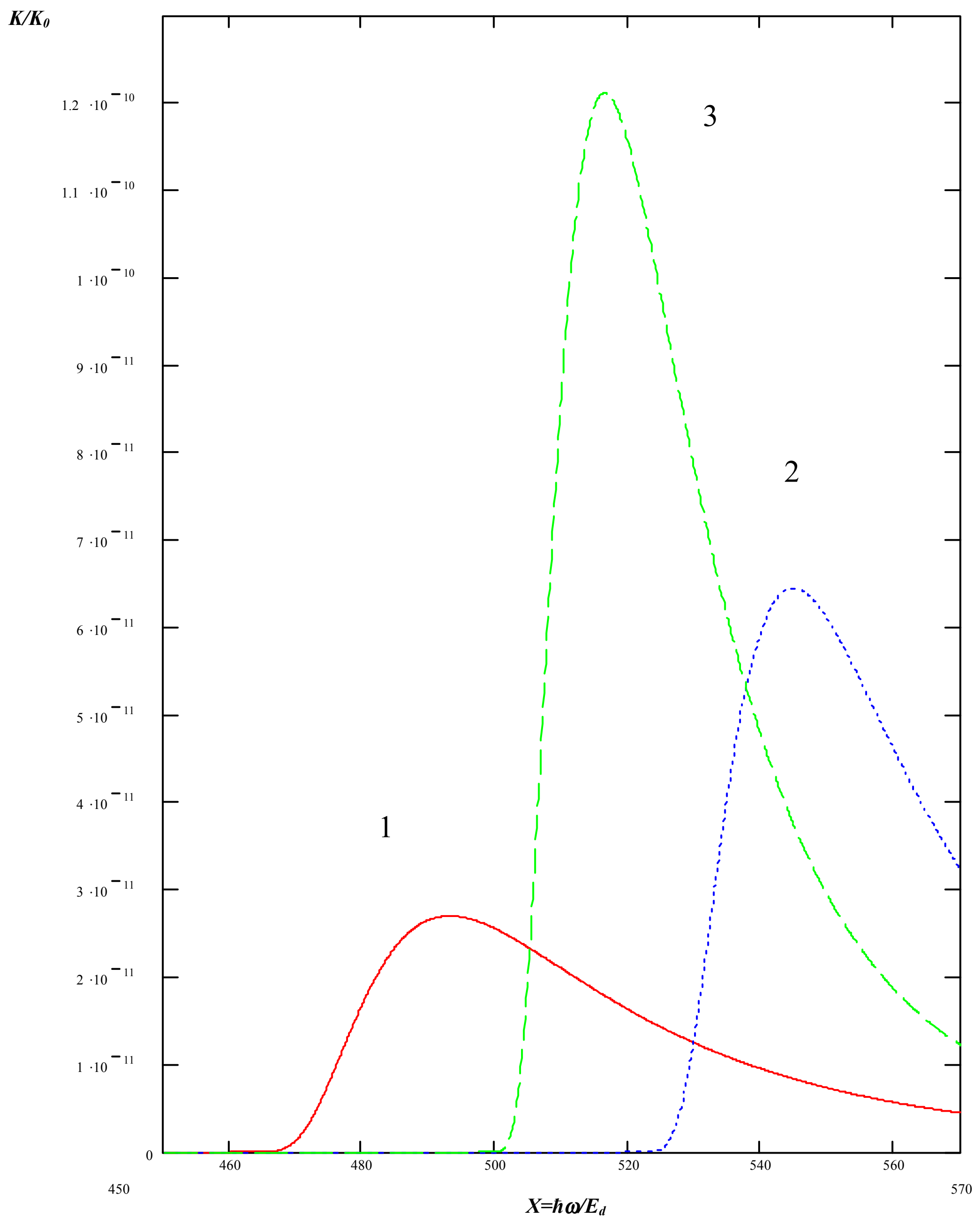

Fig. 1. The spectral dependence of normalized light impurity absorption coefficient $K/K_0$ in magnetic field for the optical transition with maximal oscillator force $(n_1=0;\ n_2=1)$ at different values of the parameters $a^*=\sqrt{\hbar/(a_d^{\,2}B|e|)}$ and $U_0^*=U_0/E_d$ $(R_0^*=1, \eta_i^2=324)$; (1) $a^*$=10, $U_0^*$=400; (2) $a^*$=0.1, $U_0^*$=400; (3) $a^*$=0.1, $U_0^*$=250.

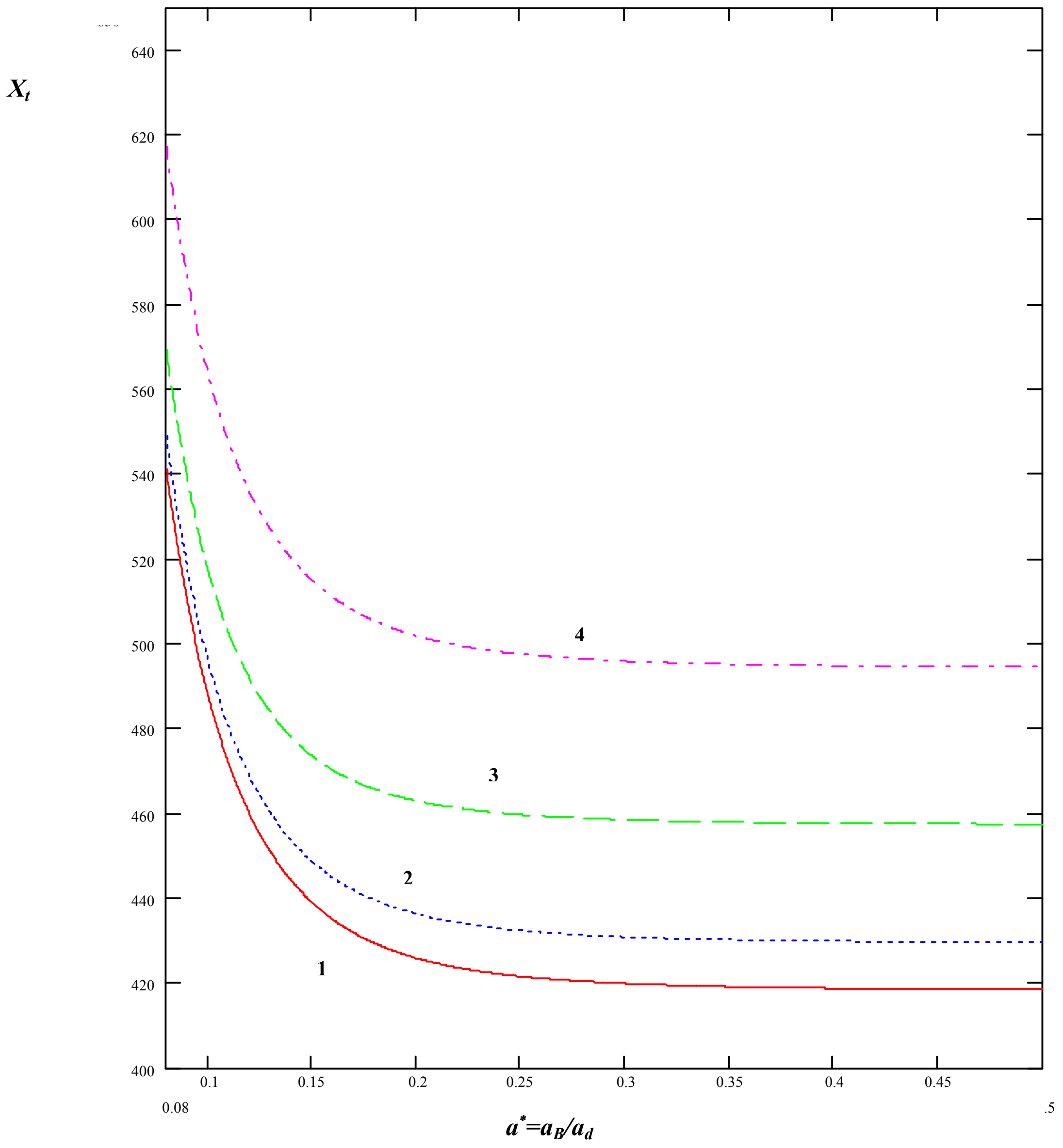

Fig. 2. The dependence of the impurity absorption threshold $X_t=(\hbar\omega)_{threshold}\,/\,E_d$ on $a^*$ for different values of QD parameters ($\overline{R_0^*}=1$): (1) $\eta_i^2$=324, $U_0^*$=200; (2) $\eta_i^2$=324, $U_0^*$=250; (3) $\eta_i^2$=324, $U_0^*$=400; (4) $\eta_i^2$=400, $U_0^*$=200.